%% file: Proceeding_ICRC_2025_ArXiv.tex
\documentclass[a4paper,11pt]{article}
\usepackage{wrapfig,pos,enumitem}
\usepackage{lineno}

\title{Scaler data from the Pierre Auger Observatory as a proxy of solar activity}
\ShortTitle{Scaler rates from the Pierre Auger Observatory: a new proxy of solar activity}

\author{C. Taricco$^{a,*}$ on behalf of the Pierre Auger Collaboration$^b$}
\author[a]{I. Bizzarri}
\author[a]{C. Dionese}
\author[c]{S. Mancuso}
\affiliation[a]{Dipartimento di Fisica, Universit\`a degli Studi di Torino, Via Pietro Giuria 1, Torino, Italy}
\affiliation[b]{Observatorio Pierre Auger, Av.\ San Mart{\'\i}n Norte 304, 5613 Malarg\"ue, Argentina\\
Full author list: {\rm\url{https://www.auger.org/archive/authors_icrc_2025.html}}}
\affiliation[c]{INAF, Osservatorio Astrofisico di Torino, via Osservatorio 20, Pino Torinese 10025, Italy} 

\emailAdd{spokespersons@auger.org}

\abstract{Solar activity variations strongly impact the modulation of the flux of low-energy Galactic  Cosmic Rays (GCRs) reaching the Earth. The secondary particles, which originate from the interaction of GCRs with the atmosphere, can be revealed by an array of ground detectors. 
We show that the low-threshold rate (scaler) time series recorded over 16 years of operation by the surface detectors of the Pierre Auger Observatory in Malargüe (Argentina) strongly  reflects solar activity and can be considered as a new proxy of solar variability. To achieve 
this result, we apply advanced spectral methods to this time series and to the classical solar sunspot number and sunspot area series. We detect and compare highly significant variations with periods ranging from the decadal to the daily scale and identify the origin of each variability mode. In conclusion, we show that the Auger scaler data, thanks to the very low noise level and high statistical significance related to the very  high count rates ($\sim 10^6$ counts per second), allow for a thorough and detailed investigation of the GCR flux variations in the heliosphere. }

\ConferenceLogo{icrc2025_16x9}

\FullConference{%
39th International Cosmic Ray Conference (ICRC2025)\\
15 -- 24 July, 2025\\
Geneva, Switzerland}

\begin{document}
\maketitle

\section{Introduction}

The propagation of Galactic Cosmic Rays (GCRs) through the heliosphere is influenced by interactions with the solar wind and the heliospheric magnetic field (HMF), which modulate their energy spectra. 
Variations in the solar activity and transient events, such as interplanetary coronal mass ejections (ICMEs) and stream interaction regions (SIRs), modify the interplanetary medium, leading to changes in the trajectories and flux of GCRs reaching Earth's atmosphere. 
This modulation occurs through a combination of diffusion, convection, and adiabatic energy losses, as described by \cite{parker1965}. 
The intensity of GCRs reaching Earth is highest during solar minimum and lowest during solar
maximum, resulting in a periodic modulation that follows the solar cycle. 
The Pierre Auger Observatory \cite{pierre2015}, located in Argentina, is the world's largest facility for studying ultra-high-energy cosmic rays above $3\times 10^{17}$ eV. Since 2005, the water-Cherenkov detectors at the Pierre Auger Observatory have also been
operated in scaler mode, recording low-threshold count rates. This additional mode enables the
detection of both transient phenomena—such as gamma-ray bursts, solar flares, and Forbush decreases—and long-term trends in GCRs modulation.  
The scaler data, characterized by high count rates (${\sim}10^6$ counts per second) and low noise, provide a detailed view of GCR flux variations on timescales ranging from decadal to daily.
In this study, we analyze a 16-year-long scaler time series, applying an Auto-Regressive (AR)  technique to fill data gaps. 
Through advanced spectral analysis, we identify significant oscillatory components and investigate their origins, particularly their relationship with solar modulation. 
The results reveal GCR variations on multiple time scales, offering insights into the underlying physical processes and highlighting the complementary role of Auger scaler data in understanding GCR dynamics in the heliosphere.

\section{Scaler rate at the Pierre Auger Observatory}

The Pierre Auger Observatory, located at 1400 meters above sea level near the town of Malarg\"ue, Argentina, combines surface detectors (SDs) with fluorescence telescopes (FDs). 
The SD array consists of 1600 water-Cherenkov detectors spaced across 3000 km², with 60 additional units in a denser infill region to extend the energy range down to lower energies \citep{allekotte2008}. 
Each detector measures Cherenkov radiation produced by cosmic-ray-induced air showers, using photomultiplier tubes (PMTs) to measure the light signals. 
Since 2005, the Observatory has also been operated in scaler mode, recording low energy signals (15 to 100 MeV) at a rate of $\sim 3 \times 10^6$ counts per second, mainly from cosmic rays between 10 GeV and a few TeV.
These data, collected every second, are affected by atmospheric conditions, instrumental instabilities, and natural fluctuations in the low-energy flux. 
To ensure high-quality data, corrections are applied for atmospheric pressure, detector ageing, and baseline drift \citep{pierre2011, Schimassek:2020soa}. 
The corrected scaler rate is normalized to a reference value from 2013, which is roughly in the middle of the dataset, ensuring consistency across detectors. 

\begin{figure*}
\centering
\includegraphics[width=14cm]{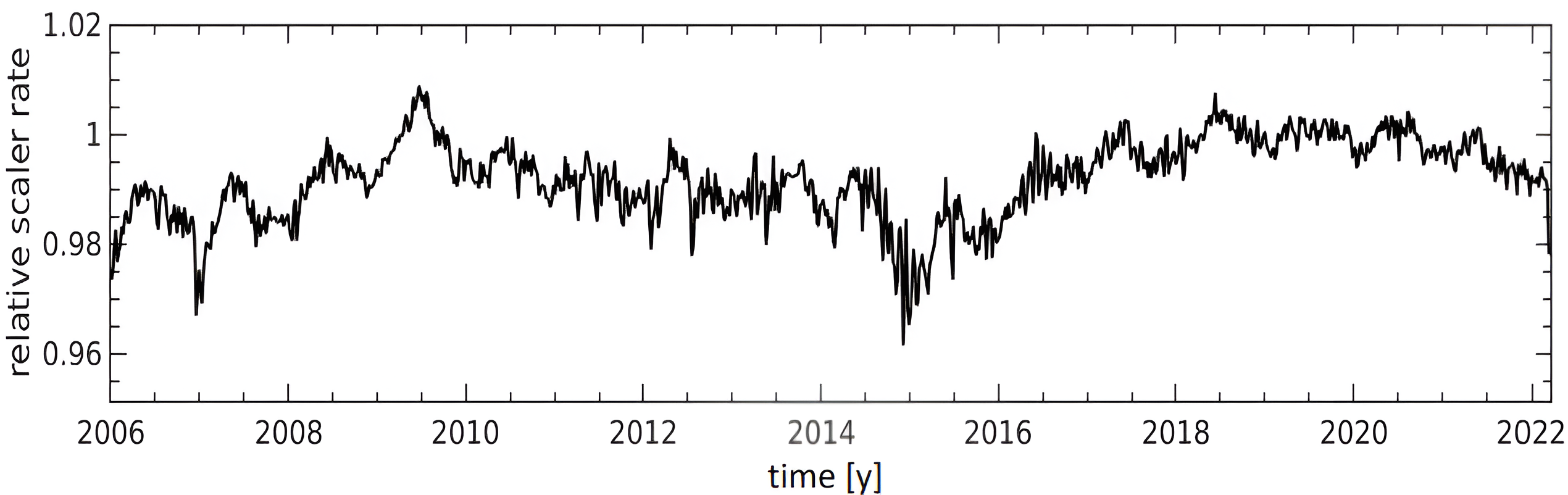}
\caption{Relative scaler rates series from 01 January 2006 to 19 March 2022. The scaler rate in this figure was obtained by resampling  the original series every 6 days after applying a gap-filling process relying on an Auto-Regressive model to the series.
The rate incorporates all the corrections detailed in ~\cite{Schimassek:2020soa} and the text.}
\label{fig:series}
\end{figure*}

Figure \ref{fig:series} shows the time series resulting from the mean of the relative scaler rate $r_i(t)$ over all stations and from the application of a gap-filling technique using an AR models to address data gaps. It results in a uniformly 6-day-sampled time series from January 2006 to March 2022. 

By visual inspection, this series reveals decadal and annual modulations in cosmic-ray flux, providing insights into solar activity and heliospheric dynamics. 

\section{Spectral analysis and results}

To accurately identify significant periodic components in the relative scaler rate series, we applied advanced spectral methods, particularly Singular Spectrum Analysis (SSA; \citep{vautard1989,vautard1992,ghil2002}). 
Unlike traditional Fourier methods, SSA uses data-adaptive basis functions, making it highly effective for analyzing short and noisy time series. 
It decomposes a time series into statistically independent components, distinguishing oscillatory patterns from noise, and can detect modulations in both amplitude and phase. 
To extract significant components from the background noise, we used a Monte Carlo-based SSA (MC-SSA) approach, which ensures robustness against statistical fluctuations inherent to the signal or measurement process.
 
\begin{figure*}
\centering
\includegraphics[width=14cm]{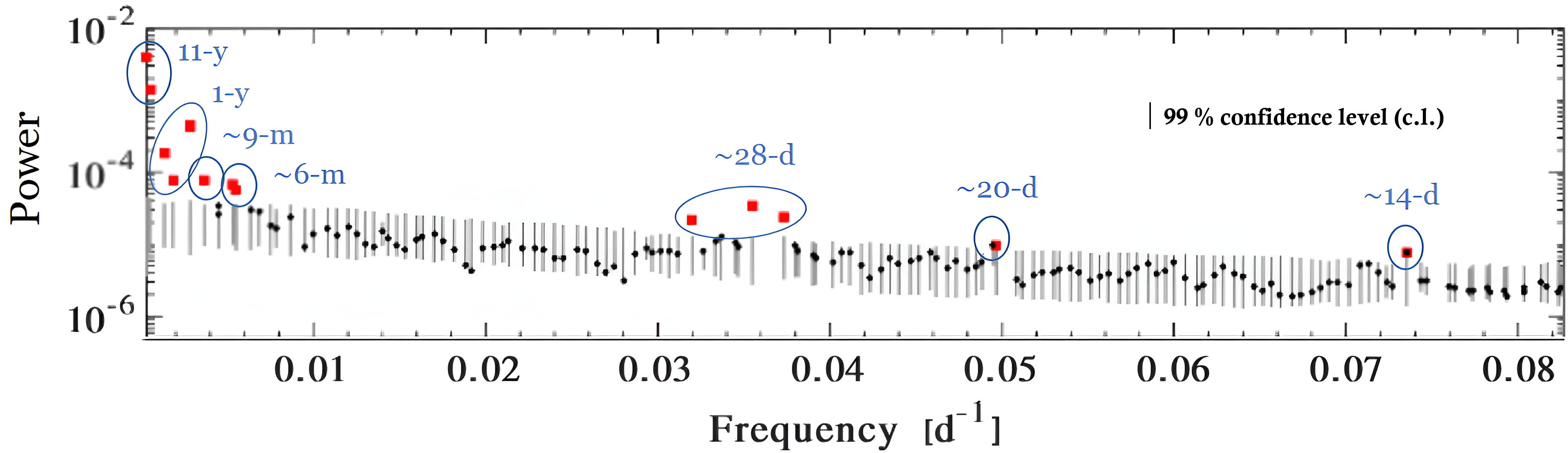}
\caption{MC-SSA spectrum of the relative scaler rate.
The Monte Carlo ensemble size is 10\,000.
The gray bars, which bracket 99\% of the power values obtained from the ensemble, represent the Monte Carlo band.
The significant spectral components are indicated by the red squares, while the black dots represent the spectral components that can be parameterized as red noise.
The significant components with the same period specified in blue are grouped with blue boundaries.
}
\label{fig:spettroSSA}
\end{figure*}

A window length of $M=150$ samples ($\sim 2.5$ years) was chosen and the results were validated across a range of $M$ values. 
The MC-SSA spectrum (Figure \ref{fig:spettroSSA}) reveals significant periodic components at a 99\% confidence level (c.l.), including a decadal trend, an annual oscillation, and shorter-term variations with periods of approximately 9 months, 6 months, 28 days, 20 days, and 14 days. 
Together, these components explain about 88\% of the total variance, with the remaining 12\% attributed to noise, highlighting the exceptionally low noise level in the scaler data.
The gray bars in Figure \ref{fig:spettroSSA} represent the Monte Carlo confidence band.
As one can see, no anomalous power exceeds this band except those corresponding to the significant components mentioned above and highlighted by the red squares.
The black dots indicate the spectral components that can be parameterized as red noise or are not significant.
The decadal trend and the annual oscillation are the most prominent, with the latter showing clear seasonal patterns. 
Shorter periodicities, such as the 9- and 6-month oscillations, are associated to solar activity, particularly the Rieger-type periodicity \citep{rieger1984}, which has been observed in various solar indices. 
The 28-day, 20-day, and 14-day periodicities were associated with solar rotation and the distribution of active regions, with the 28-day component showing higher variability during solar maximum and the declining phase of the solar cycle.
These results are compared with the sunspot number (SN) series \citep{SILSO2021}, a well-known proxy of solar activity, which exhibits similar periodicities. 
This consistency underscores the reliability of the scaler data in capturing solar-induced modulations in cosmic-ray flux. 
The study demonstrates the effectiveness of MC-SSA in extracting subtle periodic signals from complex time series, providing valuable insights into the relationship between solar activity and cosmic ray modulation.

\begin{figure*}
\centering
\includegraphics[width=14cm]{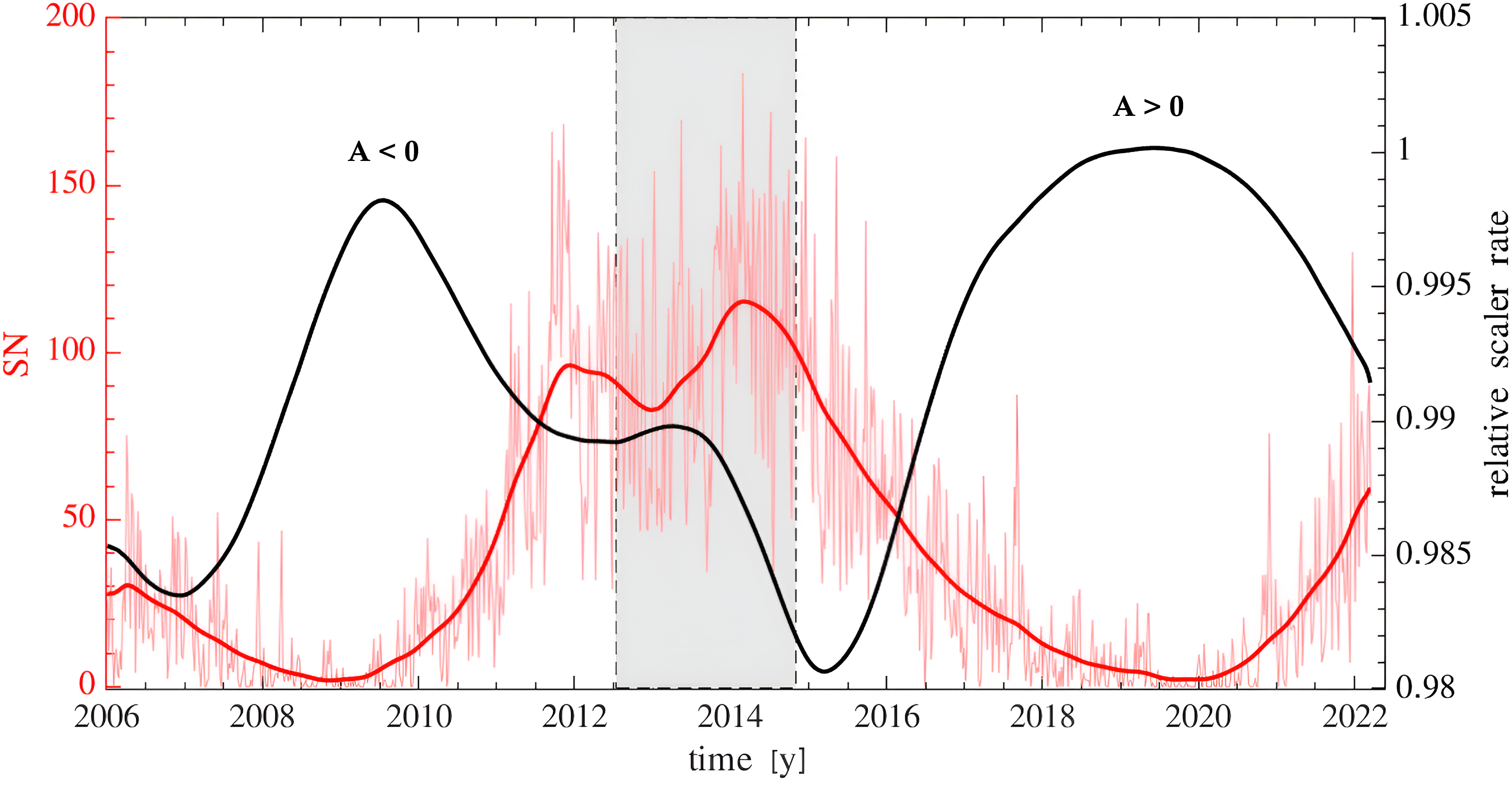}
\caption{Comparison between the decadal trend revealed in the Auger scaler rate (black curve) and the SN series sampled every 6\,d (shaded red curve), superimposed by the decadal modulation revealed in the latter by SSA (red curve).
An anticorrelation among the decadal trends is visible.
The shaded gray bar represents the total time interval required for the polar field reversal in both hemispheres from June 2012 to November 2014.}
\label{fig:confronto_decennali}
\end{figure*}

The Sun's magnetic activity, driven by a dynamo process in the convective zone, manifests cyclically through sunspots and active regions. 
The primary 11-year cycle involves the reversal of the Sun's global magnetic field polarity. 
The decadal modulation observed in the scalar data accounts for about 68\% of the signal variance and is anticorrelated with the solar decadal cycle. 
In Figure \ref{fig:confronto_decennali}, this component is directly compared to the decadal one (red curve) extracted by SSA from the SN series (light red curve) using the same window length $M$ adopted for the scaler analysis. 
However, a time lag between the decadal components of GCR intensity and solar activity is observed, with a duration varying over time. 
This lag is linked to the solar magnetic polarity reversal, affecting the drift patterns of charged particles in the heliosphere. 
During positive polarity cycles ($A>0$), protons drift inward through the polar regions, while during negative cycles ($A<0$), they drift through the equatorial regions and encounter the wavy heliospheric current sheet. 
This results in broader cosmic-ray intensity maxima during $A>0$ cycles. 
The observed phase displacement suggests a possible signature of the 22-year Hale cycle in the scaler data, although its detection is limited by the time length of the dataset.
The annual modulation in cosmic ray flux, with minima in December-January and maxima in June-July, results from both terrestrial and extraterrestrial factors: seasonal temperature variations affect atmospheric muon flux, with higher temperatures in December-January increasing muon decay rates and reducing the muon flux at ground-level; moreover, the Earth-Sun distance variation due to orbital eccentricity and the asymmetry of the heliospheric magnetic field contribute to this cycle. 
Further studies are needed to fully understand the muonic signal fraction and other atmospheric effects.

\begin{table}
\centering
\begin{tabular}{ |c| c| c| c |c|  } \hline
\multicolumn{5}{|c|}{SSA significant components (99\% c.l.)} \\
\hline
 & Scalers & Sunspot area & Sunspot area  &  Sunspot area\\ 
 &  & (total) &  (north) &   (south)\\ 
\hline
Period &  Variance [\%] & Variance [\%]& Variance [\%]& Variance [\%]\\ \hline
11 y            & 68.2 & 53.7 & 37.8 & 40.7  \\
1 y             & 14.8 &  -   &  -   &   -   \\
$ 9$ months & 1.0  & 4.0  &  -   &  6.2  \\
$ 6$ months & 1.6  & 2.4  &  3.4 (90\% c.l.)  &   -   \\
$ 28$ d     & 2.1  & 7.0  &  8.1 & 12.0  \\
$20$ d          & 0.4  & 3.0  &  3.5 &   -   \\
$ 14$ d     & 0.2  &  -   &  -   &   -   \\
\hline
\multicolumn{5}{|c|}{} \\ \hline
Signal           & $ 88\%$ & $ 70\%$ & $ 53\%$ & $ 59\%$  \\
Noise            & $ 12\%$ & $ 30\%$ & $ 47\%$ & $ 41\%$  \\
\hline 
\end{tabular}
\label{tab:tabella aree}
\caption{The percentage of the total variance associated with the SSA significant components of the three sunspot area series and the scalers series.
The last two rows show the total variance related to signal and noise for each series.}
\end{table}

Strong periodicities of approximately 6 and 9 months were identified in the scaler data. 
These components are linked to solar activity, specifically the Rieger-type periodicity, which has been observed in various solar indicators such as sunspot areas, X-ray flares, and radio flux.  
These periodicities are thought to arise from magneto-Rossby waves in the solar tachocline, which modulate the emergence of magnetic flux and, consequently, the GCR flux \citep{zaqarashvili2010}. 
A 28-day periodicity, related to solar rotation and the distribution of long-lived active regions, was detected. 
This component peaks during solar maximum and remains variable during the declining phase of the solar cycle. 
This periodicity is attributed to the quasi-rigid rotation of coronal magnetic structures \citep{mancuso2020} and corotating interaction regions (CIRs) in the solar cycle's declining phase. 
This periodicity has also been observed in neutron monitor data and solar indices \citep{,bazilevskaya2000,lopez2020}.
A 14-day oscillation was identified, showing stronger variability during the declining phases of solar cycles. 
This periodicity is associated with the occurrence of two high-speed solar wind streams approximately $180^\circ$ apart in solar longitude, resulting from the tilt of the solar dipole magnetic field. 

The spectral results obtained from the scaler series were also compared with the total sunspot area (SA). 
In Table 1, we show the percentage of total variance associated with the SSA significant components of the SA series and the scalers series.
The total SA series shows the same spectral content as the scalers series, except for the 14-day and the annual components.
The absence of the annual component in the SA series is expected, due to its terrestrial origin.

Furthermore, we performed the analysis of the two hemispheric SA series, which reveals that the 6-month periodicity is associated with the Northern Hemisphere, while the 9-month periodicity is linked to the Southern Hemisphere.

\begin{figure}
    \centering
    \includegraphics[width=14cm]{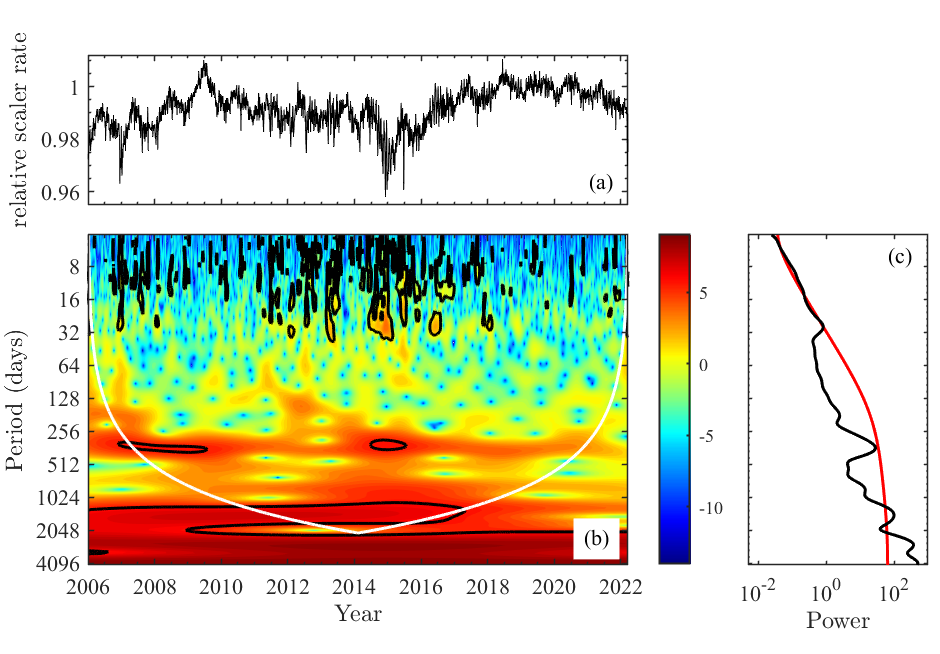}
    \caption{Continuous Wavelet power spectrum (panel b) of the Auger scaler rate sampled every 48 hours (panel a) and Global Wavelet Spectrum (panel c). The black curves in panel b enclose regions with a confidence level greater than 90\% against a red-noise process null hypothesis, while the COI, indicating regions influenced by edge effects, is represented by the white curve.}
    \label{fig:cwt_auger}
\end{figure}

All the significant SSA-detected components were also identified by the continuous wavelet transform (CWT) method.
Figure \ref{fig:cwt_auger}b shows the continuous wavelet spectrum of the scaler series (panel a), where areas with a high (low) power are represented by the color red (blue). 
The black contours enclose regions with a c.l. greater than 90\% against the null hypothesis of a red-noise process.
The cone of influence (COI), delimiting regions influenced by edge effects, is represented by the white curve.
Figure \ref{fig:cwt_auger}c shows the Global Wavelet Spectrum (GWS) (black curve), obtained by averaging the CWT spectrum over time, along with the corresponding significance levels (red curve).
The decadal modulation shows the strongest power, although it lies outside the COI due to the limited length of the series.
The annual periodicity is also detected, with increased and statistically significant power during intervals when its amplitude is higher.
The monthly oscillation is highly significant between 2012 and 2017, while it shows a lower power elsewhere. 
This result is in agreement with that found by the SSA.
Furthermore, a peak in the GWS between 8 and 16 days of period is observed, corresponding to the 14-day significant component previously discussed.
A peak at a period of 186 days is observed in the GWS (Figure \ref{fig:cwt_auger}c), although it is not significant due to the limitations of the method in detecting peaks showing a large difference in amplitude and close to each other. 

\section{Discussion and Conclusions}

This study demonstrates the significant impact of solar activity on the modulation of low-energy GCR fluxes reaching Earth, as inferred from the analysis of a 16-year time series of scaler rates measured by the Pierre Auger Observatory.
Through advanced spectral analysis, we identified several periodic components with high confidence (99\% c.l.) against red noise. 
The dominant decadal modulation is anticorrelated with the 11-year solar cycle. 
An annual oscillation, peaking during austral winters and dipping in summers, is attributed to factors like temperature effects on muon flux, Earth-Sun distance variations, and heliospheric magnetic field asymmetry.
Shorter-term oscillations with periods of about 9 and 6 months, 28, 20, and 14 days were also detected. 
The 28-day periodicity, linked to solar rotation and active regions, shows higher variability during Solar Cycle 24's maximum and declining phases. 
The 14-day component, associated with solar active longitudes and tilted dipole structures, becomes more pronounced during solar cycle declines.
The study underscores the low noise and high statistical significance of Auger scaler data, which enable a detailed analysis of GCR flux variations across timescales. 
With the implementation of AugerPrime \citep{aab2016} SD electronics, extending data availability beyond 2022, these investigations can be further extended. 
Overall, the results position Auger scaler data as a powerful tool for probing heliospheric influences on GCR modulation.

\small

\bibliography{biblio}
\bibliographystyle{JHEP}

\newpage

\par\noindent
\textbf{The Pierre Auger Collaboration}\\

\begin{wrapfigure}[8]{l}{0.12\linewidth}
\vspace{-2.9ex}
\includegraphics[width=0.98\linewidth]{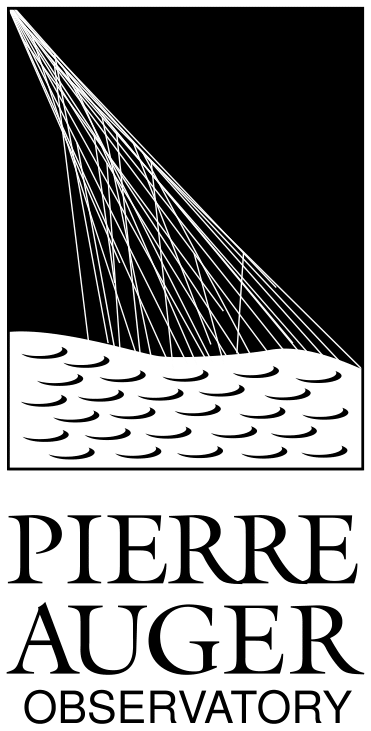}
\end{wrapfigure}
\begin{sloppypar}\noindent
\input{latex_authorlist_authors_2025-06}
\end{sloppypar}
\input{latex_authorlist_institutions_2025-06}
\input{acknowledgments_2025-06}
The SSA analyses were performed using the SSA-MTM Toolkit.\footnote{Freeware SSA-MTM Toolkit at \url{https://research.aos.ucla.edu/dkondras/ssa/form.html}.}
Daily mean sunspot numbers come from the source: WDC-SILSO, Royal Observatory of Belgium, Brussels, and these can be downloaded from \url{https://www.sidc.be/silso/}.

\end{document}

%% file: latex_authorlist_authors_2025-06.tex
A.~Abdul Halim$^{13}$,
P.~Abreu$^{70}$,
M.~Aglietta$^{53,51}$,
I.~Allekotte$^{1}$,
K.~Almeida Cheminant$^{78,77}$,
A.~Almela$^{7,12}$,
R.~Aloisio$^{44,45}$,
J.~Alvarez-Mu\~niz$^{76}$,
A.~Ambrosone$^{44}$,
J.~Ammerman Yebra$^{76}$
% ...

%% file: latex_authorlist_institutions_2025-06.tex
\begin{description}[labelsep=0.2em,align=right,labelwidth=0.7em,labelindent=0em,leftmargin=2em,noitemsep,before={\renewcommand\makelabel[1]{##1 }}]
\item[$^{1}$] Centro At\'omico Bariloche and Instituto Balseiro (CNEA-UNCuyo-CONICET), San Carlos de Bariloche, Argentina
\item[$^{2}$] Departamento de F\'\i{}sica and Departamento de Ciencias de la Atm\'osfera y los Oc\'eanos, FCEyN, Universidad de Buenos Aires and CONICET, Buenos Aires, Argentina
\item[$^{3}$] IFLP, Universidad Nacional de La Plata and CONICET, La Plata, Argentina
\item[$^{4}$] Instituto de Astronom\'\i{}a y F\'\i{}sica del Espacio (IAFE, CONICET-UBA), Buenos Aires, Argentina
\item[$^{5}$] Instituto de F\'\i{}sica de Rosario (IFIR) -- CONICET/U.N.R.\ and Facultad de Ciencias Bioqu\'\i{}micas y Farmac\'euticas U.N.R., Rosario, Argentina
\item[$^{6}$] Instituto de Tecnolog\'\i{}as en Detecci\'on y Astropart\'\i{}culas (CNEA, CONICET, UNSAM), and Universidad Tecnol\'ogica Nacional -- Facultad Regional Mendoza (CONICET/CNEA), Mendoza, Argentina
\item[$^{7}$] Instituto de Tecnolog\'\i{}as en Detecci\'on y Astropart\'\i{}culas (CNEA, CONICET, UNSAM), Buenos Aires, Argentina
\item[$^{8}$] International Center of Advanced Studies and Instituto de Ciencias F\'\i{}sicas, ECyT-UNSAM and CONICET, Campus Miguelete -- San Mart\'\i{}n, Buenos Aires, Argentina
\item[$^{9}$] Laboratorio Atm\'osfera -- Departamento de Investigaciones en L\'aseres y sus Aplicaciones -- UNIDEF (CITEDEF-CONICET), Argentina
\item[$^{10}$] Observatorio Pierre Auger, Malarg\"ue, Argentina
\item[$^{11}$] Observatorio Pierre Auger and Comisi\'on Nacional de Energ\'\i{}a At\'omica, Malarg\"ue, Argentina
\item[$^{12}$] Universidad Tecnol\'ogica Nacional -- Facultad Regional Buenos Aires, Buenos Aires, Argentina
\item[$^{13}$] University of Adelaide, Adelaide, S.A., Australia
\item[$^{14}$] Universit\'e Libre de Bruxelles (ULB), Brussels, Belgium
\item[$^{15}$] Vrije Universiteit Brussels, Brussels, Belgium
\item[$^{16}$] Centro Brasileiro de Pesquisas Fisicas, Rio de Janeiro, RJ, Brazil
\item[$^{17}$] Centro Federal de Educa\c{c}\~ao Tecnol\'ogica Celso Suckow da Fonseca, Petropolis, Brazil
\item[$^{18}$] Instituto Federal de Educa\c{c}\~ao, Ci\^encia e Tecnologia do Rio de Janeiro (IFRJ), Brazil
\item[$^{19}$] Universidade de S\~ao Paulo, Escola de Engenharia de Lorena, Lorena, SP, Brazil
\item[$^{20}$] Universidade de S\~ao Paulo, Instituto de F\'\i{}sica de S\~ao Carlos, S\~ao Carlos, SP, Brazil
\item[$^{21}$] Universidade de S\~ao Paulo, Instituto de F\'\i{}sica, S\~ao Paulo, SP, Brazil
\item[$^{22}$] Universidade Estadual de Campinas (UNICAMP), IFGW, Campinas, SP, Brazil
\item[$^{23}$] Universidade Estadual de Feira de Santana, Feira de Santana, Brazil
\item[$^{24}$] Universidade Federal de Campina Grande, Centro de Ciencias e Tecnologia, Campina Grande, Brazil
\item[$^{25}$] Universidade Federal do ABC, Santo Andr\'e, SP, Brazil
\item[$^{26}$] Universidade Federal do Paran\'a, Setor Palotina, Palotina, Brazil
\item[$^{27}$] Universidade Federal do Rio de Janeiro, Instituto de F\'\i{}sica, Rio de Janeiro, RJ, Brazil
\item[$^{28}$] Universidad de Medell\'\i{}n, Medell\'\i{}n, Colombia
\item[$^{29}$] Universidad Industrial de Santander, Bucaramanga, Colombia
\item[$^{30}$] Charles University, Faculty of Mathematics and Physics, Institute of Particle and Nuclear Physics, Prague, Czech Republic
\item[$^{31}$] Institute of Physics of the Czech Academy of Sciences, Prague, Czech Republic
\item[$^{32}$] Palacky University, Olomouc, Czech Republic
\item[$^{33}$] CNRS/IN2P3, IJCLab, Universit\'e Paris-Saclay, Orsay, France
\item[$^{34}$] Laboratoire de Physique Nucl\'eaire et de Hautes Energies (LPNHE), Sorbonne Universit\'e, Universit\'e de Paris, CNRS-IN2P3, Paris, France
\item[$^{35}$] Univ.\ Grenoble Alpes, CNRS, Grenoble Institute of Engineering Univ.\ Grenoble Alpes, LPSC-IN2P3, 38000 Grenoble, France
\item[$^{36}$] Universit\'e Paris-Saclay, CNRS/IN2P3, IJCLab, Orsay, France
\item[$^{37}$] Bergische Universit\"at Wuppertal, Department of Physics, Wuppertal, Germany
\item[$^{38}$] Karlsruhe Institute of Technology (KIT), Institute for Experimental Particle Physics, Karlsruhe, Germany
\item[$^{39}$] Karlsruhe Institute of Technology (KIT), Institut f\"ur Prozessdatenverarbeitung und Elektronik, Karlsruhe, Germany
\item[$^{40}$] Karlsruhe Institute of Technology (KIT), Institute for Astroparticle Physics, Karlsruhe, Germany
\item[$^{41}$] RWTH Aachen University, III.\ Physikalisches Institut A, Aachen, Germany
\item[$^{42}$] Universit\"at Hamburg, II.\ Institut f\"ur Theoretische Physik, Hamburg, Germany
\item[$^{43}$] Universit\"at Siegen, Department Physik -- Experimentelle Teilchenphysik, Siegen, Germany
\item[$^{44}$] Gran Sasso Science Institute, L'Aquila, Italy
\item[$^{45}$] INFN Laboratori Nazionali del Gran Sasso, Assergi (L'Aquila), Italy
\item[$^{46}$] INFN, Sezione di Catania, Catania, Italy
\item[$^{47}$] INFN, Sezione di Lecce, Lecce, Italy
\item[$^{48}$] INFN, Sezione di Milano, Milano, Italy
\item[$^{49}$] INFN, Sezione di Napoli, Napoli, Italy
\item[$^{50}$] INFN, Sezione di Roma ``Tor Vergata'', Roma, Italy
\item[$^{51}$] INFN, Sezione di Torino, Torino, Italy
\item[$^{52}$] Istituto di Astrofisica Spaziale e Fisica Cosmica di Palermo (INAF), Palermo, Italy
\item[$^{53}$] Osservatorio Astrofisico di Torino (INAF), Torino, Italy
\item[$^{54}$] Politecnico di Milano, Dipartimento di Scienze e Tecnologie Aerospaziali , Milano, Italy
\item[$^{55}$] Universit\`a del Salento, Dipartimento di Matematica e Fisica ``E.\ De Giorgi'', Lecce, Italy
\item[$^{56}$] Universit\`a dell'Aquila, Dipartimento di Scienze Fisiche e Chimiche, L'Aquila, Italy
\item[$^{57}$] Universit\`a di Catania, Dipartimento di Fisica e Astronomia ``Ettore Majorana``, Catania, Italy
\item[$^{58}$] Universit\`a di Milano, Dipartimento di Fisica, Milano, Italy
\item[$^{59}$] Universit\`a di Napoli ``Federico II'', Dipartimento di Fisica ``Ettore Pancini'', Napoli, Italy
\item[$^{60}$] Universit\`a di Palermo, Dipartimento di Fisica e Chimica ''E.\ Segr\`e'', Palermo, Italy
\item[$^{61}$] Universit\`a di Roma ``Tor Vergata'', Dipartimento di Fisica, Roma, Italy
\item[$^{62}$] Universit\`a Torino, Dipartimento di Fisica, Torino, Italy
\item[$^{63}$] Benem\'erita Universidad Aut\'onoma de Puebla, Puebla, M\'exico
\item[$^{64}$] Unidad Profesional Interdisciplinaria en Ingenier\'\i{}a y Tecnolog\'\i{}as Avanzadas del Instituto Polit\'ecnico Nacional (UPIITA-IPN), M\'exico, D.F., M\'exico
\item[$^{65}$] Universidad Aut\'onoma de Chiapas, Tuxtla Guti\'errez, Chiapas, M\'exico
\item[$^{66}$] Universidad Michoacana de San Nicol\'as de Hidalgo, Morelia, Michoac\'an, M\'exico
\item[$^{67}$] Universidad Nacional Aut\'onoma de M\'exico, M\'exico, D.F., M\'exico
\item[$^{68}$] Institute of Nuclear Physics PAN, Krakow, Poland
\item[$^{69}$] University of \L{}\'od\'z, Faculty of High-Energy Astrophysics,\L{}\'od\'z, Poland
\item[$^{70}$] Laborat\'orio de Instrumenta\c{c}\~ao e F\'\i{}sica Experimental de Part\'\i{}culas -- LIP and Instituto Superior T\'ecnico -- IST, Universidade de Lisboa -- UL, Lisboa, Portugal
\item[$^{71}$] ``Horia Hulubei'' National Institute for Physics and Nuclear Engineering, Bucharest-Magurele, Romania
\item[$^{72}$] Institute of Space Science, Bucharest-Magurele, Romania
\item[$^{73}$] Center for Astrophysics and Cosmology (CAC), University of Nova Gorica, Nova Gorica, Slovenia
\item[$^{74}$] Experimental Particle Physics Department, J.\ Stefan Institute, Ljubljana, Slovenia
\item[$^{75}$] Universidad de Granada and C.A.F.P.E., Granada, Spain
\item[$^{76}$] Instituto Galego de F\'\i{}sica de Altas Enerx\'\i{}as (IGFAE), Universidade de Santiago de Compostela, Santiago de Compostela, Spain
\item[$^{77}$] IMAPP, Radboud University Nijmegen, Nijmegen, The Netherlands
\item[$^{78}$] Nationaal Instituut voor Kernfysica en Hoge Energie Fysica (NIKHEF), Science Park, Amsterdam, The Netherlands
\item[$^{79}$] Stichting Astronomisch Onderzoek in Nederland (ASTRON), Dwingeloo, The Netherlands
\item[$^{80}$] Universiteit van Amsterdam, Faculty of Science, Amsterdam, The Netherlands
\item[$^{81}$] Case Western Reserve University, Cleveland, OH, USA
\item[$^{82}$] Colorado School of Mines, Golden, CO, USA
\item[$^{83}$] Department of Physics and Astronomy, Lehman College, City University of New York, Bronx, NY, USA
\item[$^{84}$] Michigan Technological University, Houghton, MI, USA
\item[$^{85}$] New York University, New York, NY, USA
\item[$^{86}$] University of Chicago, Enrico Fermi Institute, Chicago, IL, USA
\item[$^{87}$] University of Delaware, Department of Physics and Astronomy, Bartol Research Institute, Newark, DE, USA
\item[] -----
\item[$^{a}$] Max-Planck-Institut f\"ur Radioastronomie, Bonn, Germany
\item[$^{b}$] also at Kapteyn Institute, University of Groningen, Groningen, The Netherlands
\item[$^{c}$] School of Physics and Astronomy, University of Leeds, Leeds, United Kingdom
\item[$^{d}$] Fermi National Accelerator Laboratory, Fermilab, Batavia, IL, USA
\item[$^{e}$] Pennsylvania State University, University Park, PA, USA
\item[$^{f}$] Colorado State University, Fort Collins, CO, USA
\item[$^{g}$] Louisiana State University, Baton Rouge, LA, USA
\item[$^{h}$] now at Graduate School of Science, Osaka Metropolitan University, Osaka, Japan
\item[$^{i}$] Institut universitaire de France (IUF), France
\item[$^{j}$] now at Technische Universit\"at Dortmund and Ruhr-Universit\"at Bochum, Dortmund and Bochum, Germany
\end{description}

%% file: acknowledgments_2025-06.tex
\section*{Acknowledgments}

\begin{sloppypar}
The successful installation, commissioning, and operation of the Pierre
Auger Observatory would not have been possible without the strong
commitment and effort from the technical and administrative staff in
Malarg\"ue. We are very grateful to the following agencies and
organizations for financial support:
\end{sloppypar}

\begin{sloppypar}
Argentina -- Comisi\'on Nacional de Energ\'\i{}a At\'omica; Agencia Nacional de
Promoci\'on Cient\'\i{}fica y Tecnol\'ogica (ANPCyT); Consejo Nacional de
Investigaciones Cient\'\i{}ficas y T\'ecnicas (CONICET); Gobierno de la
Provincia de Mendoza; Municipalidad de Malarg\"ue; NDM Holdings and Valle
Las Le\~nas; in gratitude for their continuing cooperation over land
access; Australia -- the Australian Research Council; Belgium -- Fonds
de la Recherche Scientifique (FNRS); Research Foundation Flanders (FWO),
Marie Curie Action of the European Union Grant No.~101107047; Brazil --
Conselho Nacional de Desenvolvimento Cient\'\i{}fico e Tecnol\'ogico (CNPq);
Financiadora de Estudos e Projetos (FINEP); Funda\c{c}\~ao de Amparo \`a
Pesquisa do Estado de Rio de Janeiro (FAPERJ); S\~ao Paulo Research
Foundation (FAPESP) Grants No.~2019/10151-2, No.~2010/07359-6 and
No.~1999/05404-3; Minist\'erio da Ci\^encia, Tecnologia, Inova\c{c}\~oes e
Comunica\c{c}\~oes (MCTIC); Czech Republic -- GACR 24-13049S, CAS LQ100102401,
MEYS LM2023032, CZ.02.1.01/0.0/0.0/16{\textunderscore}013/0001402,
CZ.02.1.01/0.0/0.0/18{\textunderscore}046/0016010 and
CZ.02.1.01/0.0/0.0/17{\textunderscore}049/0008422 and CZ.02.01.01/00/22{\textunderscore}008/0004632;
France -- Centre de Calcul IN2P3/CNRS; Centre National de la Recherche
Scientifique (CNRS); Conseil R\'egional Ile-de-France; D\'epartement
Physique Nucl\'eaire et Corpusculaire (PNC-IN2P3/CNRS); D\'epartement
Sciences de l'Univers (SDU-INSU/CNRS); Institut Lagrange de Paris (ILP)
Grant No.~LABEX ANR-10-LABX-63 within the Investissements d'Avenir
Programme Grant No.~ANR-11-IDEX-0004-02; Germany -- Bundesministerium
f\"ur Bildung und Forschung (BMBF); Deutsche Forschungsgemeinschaft (DFG);
Finanzministerium Baden-W\"urttemberg; Helmholtz Alliance for
Astroparticle Physics (HAP); Helmholtz-Gemeinschaft Deutscher
Forschungszentren (HGF); Ministerium f\"ur Kultur und Wissenschaft des
Landes Nordrhein-Westfalen; Ministerium f\"ur Wissenschaft, Forschung und
Kunst des Landes Baden-W\"urttemberg; Italy -- Istituto Nazionale di
Fisica Nucleare (INFN); Istituto Nazionale di Astrofisica (INAF);
Ministero dell'Universit\`a e della Ricerca (MUR); CETEMPS Center of
Excellence; Ministero degli Affari Esteri (MAE), ICSC Centro Nazionale
di Ricerca in High Performance Computing, Big Data and Quantum
Computing, funded by European Union NextGenerationEU, reference code
CN{\textunderscore}00000013; M\'exico -- Consejo Nacional de Ciencia y Tecnolog\'\i{}a
(CONACYT) No.~167733; Universidad Nacional Aut\'onoma de M\'exico (UNAM);
PAPIIT DGAPA-UNAM; The Netherlands -- Ministry of Education, Culture and
Science; Netherlands Organisation for Scientific Research (NWO); Dutch
national e-infrastructure with the support of SURF Cooperative; Poland
-- Ministry of Education and Science, grants No.~DIR/WK/2018/11 and
2022/WK/12; National Science Centre, grants No.~2016/22/M/ST9/00198,
2016/23/B/ST9/01635, 2020/39/B/ST9/01398, and 2022/45/B/ST9/02163;
Portugal -- Portuguese national funds and FEDER funds within Programa
Operacional Factores de Competitividade through Funda\c{c}\~ao para a Ci\^encia
e a Tecnologia (COMPETE); Romania -- Ministry of Research, Innovation
and Digitization, CNCS-UEFISCDI, contract no.~30N/2023 under Romanian
National Core Program LAPLAS VII, grant no.~PN 23 21 01 02 and project
number PN-III-P1-1.1-TE-2021-0924/TE57/2022, within PNCDI III; Slovenia
-- Slovenian Research Agency, grants P1-0031, P1-0385, I0-0033, N1-0111;
Spain -- Ministerio de Ciencia e Innovaci\'on/Agencia Estatal de
Investigaci\'on (PID2019-105544GB-I00, PID2022-140510NB-I00 and
RYC2019-027017-I), Xunta de Galicia (CIGUS Network of Research Centers,
Consolidaci\'on 2021 GRC GI-2033, ED431C-2021/22 and ED431F-2022/15),
Junta de Andaluc\'\i{}a (SOMM17/6104/UGR and P18-FR-4314), and the European
Union (Marie Sklodowska-Curie 101065027 and ERDF); USA -- Department of
Energy, Contracts No.~DE-AC02-07CH11359, No.~DE-FR02-04ER41300,
No.~DE-FG02-99ER41107 and No.~DE-SC0011689; National Science Foundation,
Grant No.~0450696, and NSF-2013199; The Grainger Foundation; Marie
Curie-IRSES/EPLANET; European Particle Physics Latin American Network;
and UNESCO.
\end{sloppypar}

%% file: Proceeding_ICRC_2025_ArXiv.bbl
\providecommand{\href}[2]{#2}\begingroup\raggedright\begin{thebibliography}{10}

\bibitem{parker1965}
E.~N. Parker, \emph{The passage of energetic charged particles through interplanetary space}, {\emph{Planetary \& Space Science} {\bfseries 13} (1965) 9}.

\bibitem{pierre2015}
{Pierre Auger Collaboration}, \emph{{The Pierre Auger cosmic ray observatory}}, {\emph{Nuclear Instruments and Methods in Physics Research Section A: Accelerators, Spectrometers, Detectors and Associated Equipment} {\bfseries 798} (2015) 172}.

\bibitem{allekotte2008}
I.~Allekotte, A.~Barbosa, P.~Bauleo, C.~Bonifazi, B.~Civit, C.~Escobar et~al., \emph{The surface detector system of the pierre auger observatory}, {\emph{Nuclear Instruments and Methods in Physics Research Section A: Accelerators, Spectrometers, Detectors and Associated Equipment} {\bfseries 586} (2008) 409}.

\bibitem{pierre2011}
{Pierre Auger Collaboration}, \emph{{The Pierre Auger Observatory scaler mode for the study of solar activity modulation of galactic cosmic rays}}, {\emph{JINST} {\bfseries 6} (2011) 1003}.

\bibitem{Schimassek:2020soa}
{\scshape Pierre Auger} collaboration, M.~Schimassek, \emph{{Analysis of Data from the Low-Energy Modes of the Surface Detector of the Pierre Auger Observatory}},  in \emph{{Proceedings of the 32nd International Cosmic Ray Conference (ICRC 2019), 1147}}, 2020.

\bibitem{vautard1989}
R.~{Vautard} and M.~{Ghil}, \emph{{Singular spectrum analysis in nonlinear dynamics, with applications to paleoclimatic time series}}, \href{https://doi.org/10.1016/0167-2789(89)90077-8}{\emph{Physica D Nonlinear Phenomena} {\bfseries 35} (1989) 395}.

\bibitem{vautard1992}
R.~{Vautard}, P.~{Yiou} and M.~{Ghil}, \emph{{Singular-spectrum analysis: A toolkit for short, noisy chaotic signals}}, \href{https://doi.org/10.1016/0167-2789(92)90103-T}{\emph{Physica D Nonlinear Phenomena} {\bfseries 58} (1992) 95}.

\bibitem{ghil2002}
M.~{Ghil}, M.~R. {Allen}, M.~D. {Dettinger}, K.~{Ide}, D.~{Kondrashov}, M.~E. {Mann} et~al., \emph{{Advanced Spectral Methods for Climatic Time Series}}, \href{https://doi.org/10.1029/2000RG000092}{\emph{Rev. Geophys.} {\bfseries 40} (2002) 1003}.

\bibitem{rieger1984}
E.~{Rieger}, G.~H. {Share}, D.~J. {Forrest}, G.~{Kanbach}, C.~{Reppin} and E.~L. {Chupp}, \emph{{A 154-day periodicity in the occurrence of hard solar flares?}}, \href{https://doi.org/10.1038/312623a0}{\emph{Nature} {\bfseries 312} (1984) 623}.

\bibitem{SILSO2021}
{SILSO World Data Center}, \emph{{International sunspot number monthly bulletin and online catalogue}}, {\emph{Royal Observatory of Belgium Brussels} (2021) }.

\bibitem{zaqarashvili2010}
T.~V. {Zaqarashvili}, M.~{Carbonell}, R.~{Oliver} and J.~L. {Ballester}, \emph{{Magnetic Rossby Waves in the Solar Tachocline and Rieger-Type Periodicities}}, \href{https://doi.org/10.1088/0004-637X/709/2/749}{\emph{The Astrophysical Journal} {\bfseries 709} (2010) 749}.

\bibitem{mancuso2020}
S.~{Mancuso}, S.~{Giordano}, D.~{Barghini} and D.~{Telloni}, \emph{{Differential rotation of the solar corona: A new data-adaptive multiwavelength approach}}, \href{https://doi.org/10.1051/0004-6361/202039094}{\emph{Astronomy \& Astrophysics} {\bfseries 644} (2020) A18}.

\bibitem{bazilevskaya2000}
G.~A. {Bazilevskaya}, \emph{{Observations of Variability in Cosmic Rays}}, \href{https://doi.org/10.1023/A:1026721912992}{\emph{Space Science Reviews} {\bfseries 94} (2000) 25}.

\bibitem{lopez2020}
A.~{L{\'o}pez-Comazzi} and J.~J. {Blanco}, \emph{{Short-Term Periodicities Observed in Neutron Monitor Counting Rates}}, \href{https://doi.org/10.1007/s11207-020-01649-5}{\emph{Solar Physics} {\bfseries 295} (2020) 81}.

\bibitem{aab2016}
{Pierre Auger Collaboration}, \emph{The pierre auger observatory upgrade-preliminary design report}, {\emph{arXiv preprint arXiv:1604.03637} (2016) }.

\end{thebibliography}\endgroup
